\documentclass[twocolumn,showpacs,aps,prl,superscriptaddress]{revtex4}

\usepackage{graphicx}
\usepackage{dcolumn}
\usepackage{bm}
 \usepackage{amsmath}
 \usepackage{epsfig}
 \usepackage{subfigure}

 \input pubboard/babarsym
 \input Definitions

\begin{document}


 \begin{flushleft}
 \babar-PUB-05/041 \\
 SLAC-PUB-11545 \\
 hep-ex/0510070 \\
 \end{flushleft}

\title{Measurements of the Absolute Branching Fractions of \boldmath$B^\pm \to K^\pm X_{c\bar c}$}

%
\author{B.~Aubert}
\author{R.~Barate}
\author{D.~Boutigny}
\author{F.~Couderc}
\author{Y.~Karyotakis}
\author{J.~P.~Lees}
\author{V.~Poireau}
\author{V.~Tisserand}
\author{A.~Zghiche}
\affiliation{Laboratoire de Physique des Particules, F-74941 Annecy-le-Vieux, France }
\author{E.~Grauges}
\affiliation{IFAE, Universitat Autonoma de Barcelona, E-08193 Bellaterra, Barcelona, Spain }
% ...
\collaboration{The \babar\ Collaboration}
\noaffiliation

\date{\today}

\preprint{\hbox to\hsize{{\babar\ }Analysis Document \#1205, Version 06r2 PRL
draft for Final Notice\hfil}}

\begin{abstract}
We study the two-body decays of $B^\pm$ mesons to $K^\pm$ and a 
charmonium state,  $X_{c\bar c}$, in 
a sample of 210.5 fb$^{-1}$ of data from the  
\babar\  experiment. 
We perform measurements of absolute  branching fractions 
\BR($B^\pm \to K^\pm X_{c\bar c}$)  
using a missing mass technique,
and report several new or improved results.   
In particular, the upper limit \BR$(B^\pm \to
K^\pm X(3872))<3.2 \times 10^{-4}$ at 90\% CL and the inferred 
lower limit \BR$(X(3872) \to J/\psi\pi^+\pi^-)>4.2\%$ will
help in understanding the nature of the recently discovered $X(3872)$.
\end{abstract}

\maketitle

Several exclusive decays of $B$ mesons of the form $B^\pm \to K^\pm
X_{c\bar c}$ (where $X_{c\bar c}$ is one of the charmonium states \etac\!, \jpsi\!,
\chic0\!, \chicone\!, \etacp\!, \psip\!, \psipp\!), have been observed by
reconstructing the charmonium state from its decay
to some known final state, $f$\!
\cite{Aubert:2002hc,Choi:2002na}.
In principle,  
such $B$ decays provide a direct probe of charmonium properties since the
phase space is large for all known states and all should be produced
roughly equally, in the absence of a strong selection rule \cite{Quigg:2004nv}. 
However with this technique only the product of the two branching fractions
\BR$(B^\pm \to K^\pm X_{c\bar c})\times$\BR$(X_{c\bar c} \to f)$ is measured, thereby reducing 
the precision of  \BR($B^\pm \to K^\pm X_{c\bar c}$) 
when the daughter branching fraction is poorly known.  

We describe here a complementary approach, based on the measurement of the 
kaon momentum spectrum in the $B$ center-of-mass frame, where
two-body decays can be identified by  their characteristic monochromatic
line, allowing an absolute determination of \BR$(B^\pm \to K^\pm X_{c\bar
c})$. Knowledge of the $B$ center-of-mass system is obtained by exclusive
reconstruction of the other $B$ meson from a $\Upsilon (4S)$  decay. 
In addition to obtaining new information on known
charmonium states, this method is used to search for  
the $X(3872)$ state, recently observed 
in $B^\pm\to K^\pm X(3872)$ decays by Belle \cite{Choi:2003ue} 
and \babar\  \cite{Aubert:2004ns}, in the subsequent decay $X(3872) \to J/\psi \pi^+\pi^-$. 
The same method allows a search for charged partners of the $X(3872)$ in \bz
decays, independent of the $X(3872)^\pm$ decay mode. 
The nature of $X(3872)$ resonance is still unclear, 
different interpretations \cite{interpretations} have been proposed 
but more experimental data will be needed to discriminate between them.

For this analysis we use a data sample of 210.5 fb$^{-1}$ integrated 
luminosity, corresponding to $231.8 \times 10^6$ $B \bar B$ pairs. 
The data have been collected with the 
\babar\  detector at the SLAC PEP-II asymmetric-energy collider, 
where 9 GeV electrons and 3.1 GeV positrons collide at a
center-of-mass energy 10.58 GeV, corresponding to the mass of the 
$\Upsilon (4S)$ resonance.
A detailed description of the \babar\  detector can be found in 
\cite{Aubert:2001tu}. 
Charged tracks are reconstructed with a 5 layer silicon vertex 
tracker (SVT) and a 40 layer drift chamber (DCH), located in a 1.5 T
magnetic field generated by a superconducting solenoid. 
The energy of photons and electrons is
measured with an electromagnetic calorimeter 
made up of CsI(Tl) crystals.
Charged hadron identification is done with ionization measurements 
in the SVT and DCH and with an internally reflecting ring imaging Cherenkov detector. 
The instrumented flux return of the solenoid 
is used to identify muons.

The analysis is performed on a sample of events where a $B$ meson 
is fully reconstructed ($B_{recon}$). For these events, the momentum of the 
other $B$ ($B_{signal}$) can be calculated from the momentum of $B_{recon}$ 
and the beam parameters. 
We select events with a $K^\pm$ not used for the reconstruction of $B_{recon}$ and
calculate its momentum ($p_K$) in the $B_{signal}$ center of mass system.

$B_{recon}$ mesons are reconstructed in their decays to exclusive  $D^{(*)} H$ final states, where 
$H$ is one of several combinations of $\pi^\pm$, $K^\pm$, $\pi^0$ and $K_S^0$ hadrons; 
a detailed description of the method can be found in \cite{Aubert:2003zw}.

The  number of $B^\pm$ events in the data is determined with a fit to the distribution 
of the beam energy substituted mass $m_{ES} = \sqrt{E_{CM}^2/4 - p_B^2}$, where 
$E_{CM}$ is the total center-of-mass energy, determined from the beam 
parameters, and $p_B$ is the measured momentum of $B_{recon}$ 
in the center-of-mass frame. 
The fit function is the sum of a Crystal Ball function \cite{crystalball} describing the 
signal and an ARGUS function \cite{Albrecht:1993fr} for each background 
component ($e^+ e^- \to q \bar{q}$ where $q$ is $u$, $d$, $s$ or $c$ or misreconstructed $B$s),  
the relative weights of which are obtained from a Monte Carlo 
simulation (MC), 
while the total normalization factor is determined from the data.
A total of $378580 \pm 1110$ events with a fully reconstructed \bch is 
obtained.

\begin{figure}
\epsfig{figure=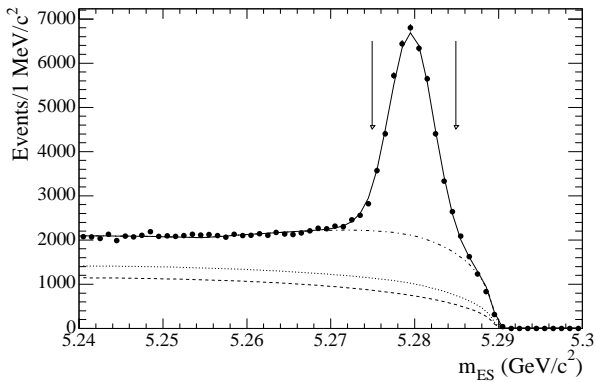,width=.45\textwidth,height=.3\textwidth}
\caption{The \mes distribution of all $B_{recon}$ after 
the NN1 selection. The solid line represents a fit described in the text;  
cumulative background contributions from $e^+ e^- \to q \bar{q}$ where $q$ is $u$, $d$, $s$ or $c$ (dashed line), \bz (dotted line), \bch (dash-dot line) 
events are shown. The arrows indicate the cuts used in the analysis (see text).}
\label{data_bchbch_fig}
\end{figure}

Fifteen variables related to the $B_{recon}$ decay characteristics, 
its production kinematics, the topology of the full event, and the angular 
correlation between $B_{recon}$ and the rest of the event are 
used in a neural network (NN1) to reduce the large background, mainly due to non-$B$ events. 
The network has 80\% signal efficiency while rejecting 90\% 
of the background. 
The \mes distribution after this selection is shown in Fig. \ref{data_bchbch_fig}.
Only events with $5.275 < m_{ES} < 5.285$ GeV/$c^2$ are used in the analysis.

We now consider only 
tracks not associated with $B_{recon}$.
Most $K^\pm$ produced in $B^\pm$ decays originate from $D$ mesons 
and their spectrum, although broad, peaks at low $p_K$.
In the $B^\pm$ rest frame, these 
$K^\pm$  are  embedded  in a ``minijet'' of $D$ decay products, 
while signal $K^\pm$ recoil against a massive (3--4 GeV/$c^2$)
state and therefore tend to be more isolated. 
A second neural network (NN2) rejects background from secondary 
$K^\pm$, 
by using  
fifteen input variables describing the energy and track
multiplicities measured 
in the $K^\pm$ hemisphere, 
the sphericity of the recoil
system, and the angular correlations between the $K^\pm$ and the recoil
system. These variables have been chosen to be independent of
the particular decay topology
of the recoil system.
Since the topology of the event changes with the recoil mass, we have considered separately two recoil 
mass regions in the training of this neural network: the ``high-mass''
region,  corresponding to 1.0$< p_K <$1.5 GeV/$c$
and the ``low-mass'' region, for 1.5$< p_K <$2.0 GeV/$c$. 
The signal training sample is $B^\pm \to K^\pm X_{c\bar c}$ MC simulation 
while the background sample consists of simulated $K^\pm$ from 
$D$ meson decays in the same momentum range.
The chosen cuts on the NN2 outputs correspond to 85\% signal efficiency; 
the background rejection factor varies between 2.5 in the $X(3872)$ and \psip region and 1.5 in 
the \jpsi region.
The selection criteria are optimized for MC signal significance with the
high-mass region blinded.

The kaon momentum distribution shows 
a series of peaks due to the two-body decays $B^\pm \to K^\pm X_{c\bar c}$ 
corresponding to the different $X_{c\bar c}$ masses,
superimposed on 
a smooth spectrum due to $K^\pm$ coming from multi-body $B^\pm$ decays,  
or non-$B^\pm$ background.
The mass of the $X_{c\bar c}$ 
state ($m_X$) can be calculated directly from $p_K$ 
using $m_X = \sqrt{m_B^2 + m_K^2 - 2 E_K m_B}$, where 
$m_B$ and $m_K$ are the $B^\pm$ and $K^\pm$ masses and $E_K$ is the 
$K^\pm$ energy.  
The resonance width $\Gamma_X$ can be obtained from the Breit-Wigner width 
of the peak in the $p_K$spectrum $\Gamma_K$, obtained after deconvolution 
with the momentum resolution function, using $\Gamma_X = \Gamma_K \beta_K m_B/m_X$, 
where $\beta_K = p_K/E_K$. 

We determine the number of $B^\pm \to K^\pm X_{c\bar c}$ events ($N_X$) 
from a fit to the $p_K$ distribution. 
The branching fraction for the decay channel is calculated as:\hfill
\vspace{-0.18cm}
\begin{equation*}
  {\cal B}(B^\pm \to K^\pm X_{c\bar c}) = \frac{N_X}{\epsilon_X \cdot N_B} , 
\end{equation*}
\noindent
where $\epsilon_X$ is the efficiency determined from the MC  
and $N_B$ the number of $B^\pm$ mesons 
in the sample. 
An alternative method, which we use to improve the branching fraction measurement 
in the case of \etac, is to normalize to the channel $B^\pm \to K^\pm J/\psi$, 
which is well-measured in the literature \cite{Eidelman:2004wy}, according to:\hfill 
\vspace{-0.18cm}
\begin{equation*}
    {\cal B}(B^\pm \to K^\pm X_{c\bar c}) = \frac{N_X}{N_{J/\psi}} \cdot 
        \frac{\epsilon_{J/\psi}}{\epsilon_X} \cdot {\cal B}(B^\pm \to K^\pm J/\psi) . 
\end{equation*}
\noindent
In this relative measurement, the systematic errors 
that are common to both resonances cancel in the ratio. 
The two methods are combined to extract ${\rm \BR}(B^\pm \to K^\pm
\eta_c)$, taking into account the correlations between them. 

\begin{figure}
\centering
\epsfig{figure=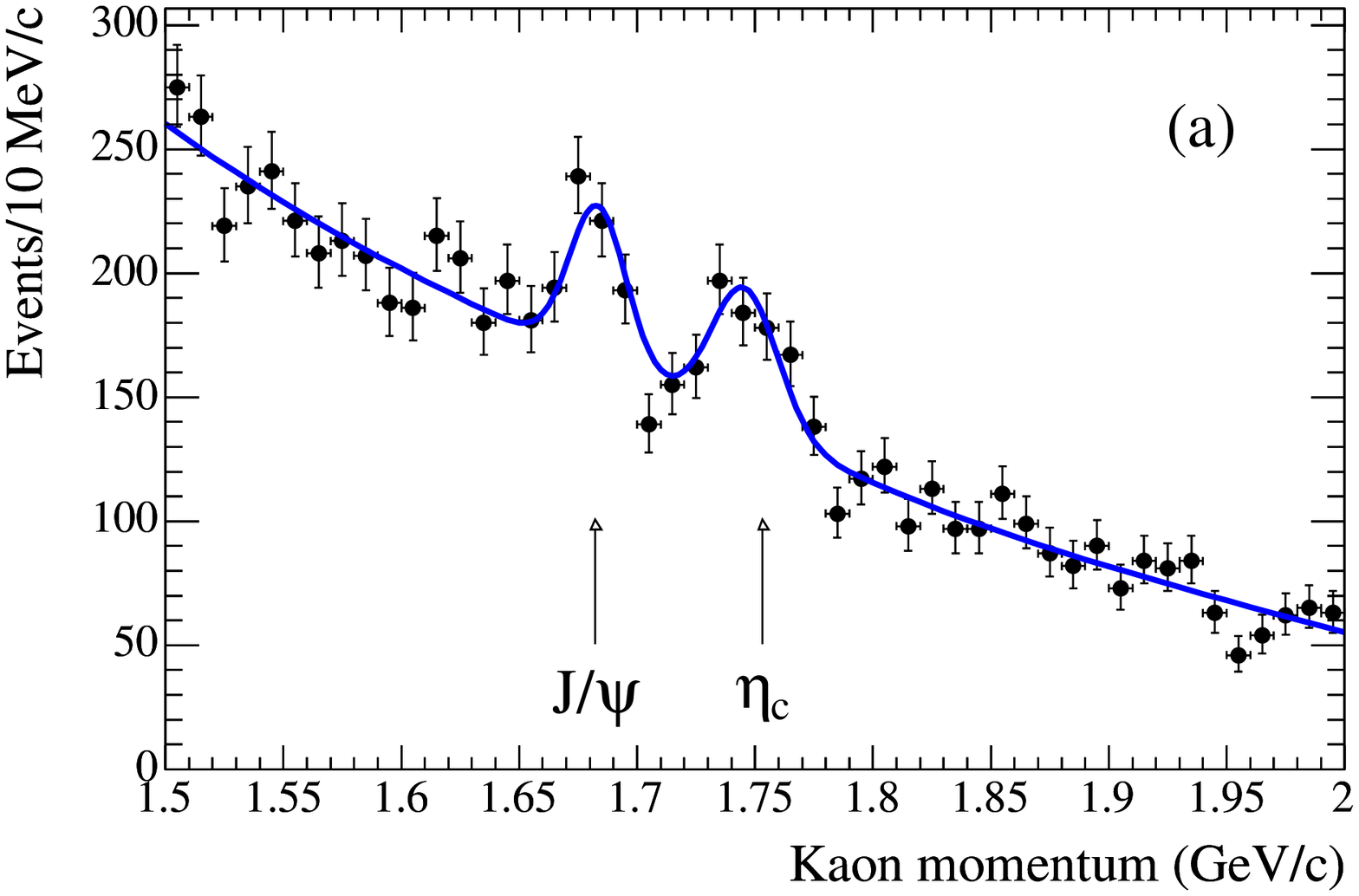,width=.45\textwidth}
\epsfig{figure=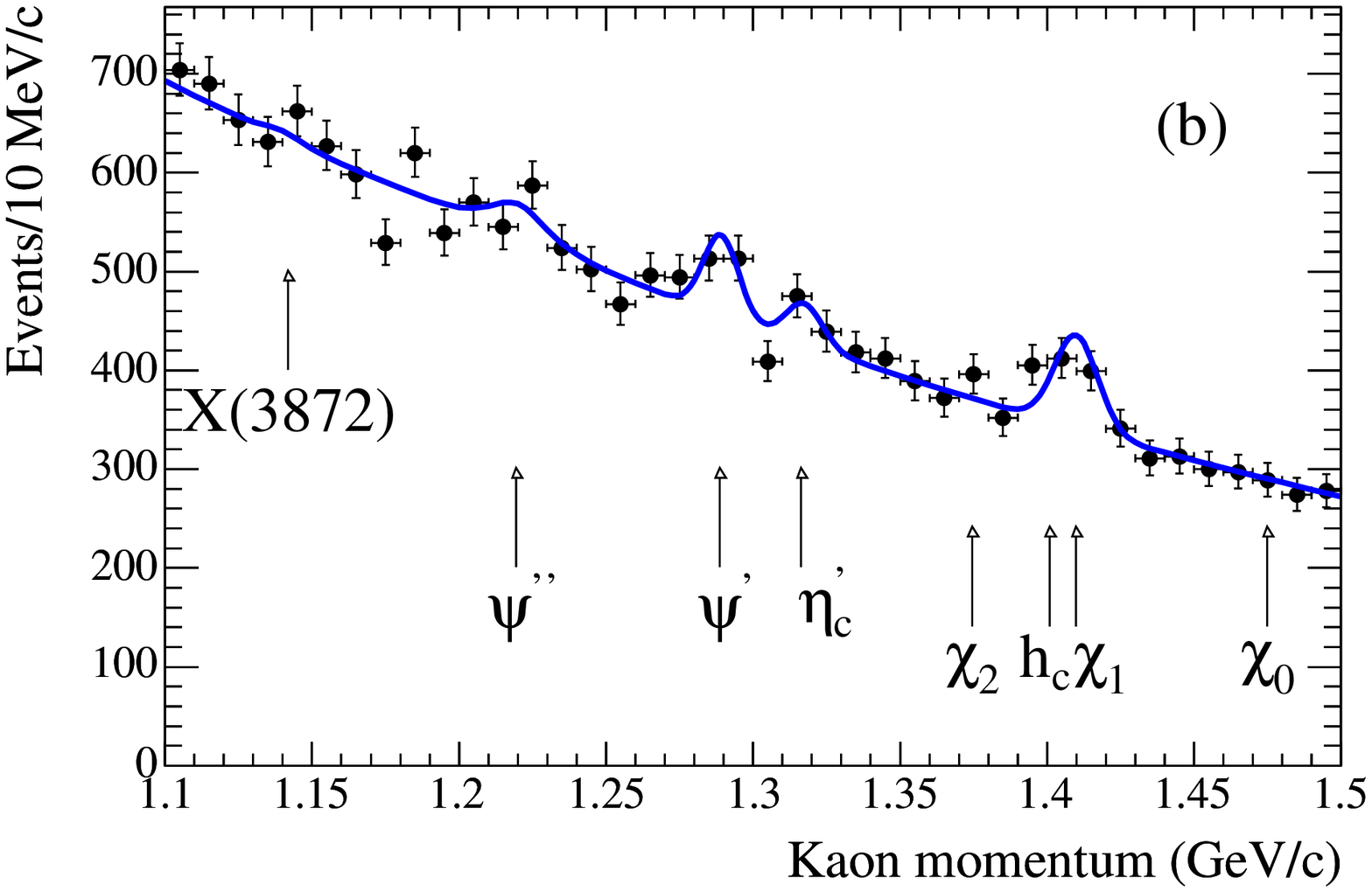,width=.45\textwidth}
\caption{Kaon momentum spectrum for the (a) low-mass and (b) high-mass regions. 
The lines represent the fit described in the text. Arrows show the 
the expected positions of known charmonium
states.}
\label{psisignal_data}
\end{figure}

We fit the $p_K$ spectrum using an unbinned maximum likelihood method.
The background is well modeled by a third degree polynomial and  
each signal is a Breit-Wigner function folded with a resolution function. 
The masses and widths of the \etac and \etacp mesons are left free;  
all others are fixed to values from reference \cite{Eidelman:2004wy}.
The resolution function has two parts: 
a Gaussian with $\sigma$ varying 
from 6 MeV/$c$ at $p_K \simeq 1.1$ GeV/$c$ 
to 12 MeV/$c$ at $p_K \simeq 1.7$ GeV/$c$  
describes the 72.5\% of the signal where $B_{recon}$ is correctly reconstructed; 
if $B_{recon}$ is incorrect, but has \mes within our range, the $p_K$
resolution is a bifurcated Gaussian with $\sigma=$ 
78 and 52 MeV/$c$ on the left and right hand side of the peak respectively.   

The spectrum in the low-mass region is expected to exhibit two peaks, 
at $p_K=1.683$ GeV/$c$ corresponding to the \jpsi, and  at $p_K=1.754$ GeV/$c$ for 
the \etac meson. 
These two  peaks are clearly seen in Fig. \ref{psisignal_data}(a);   
both have a significance of $\sim7\sigma$.
The number of events under each peak obtained from the fit is 
$N(J/\psi) = 259\pm41$ and $N(\eta_c) = 273\pm43$.

The spectrum in the high-mass region is  
fitted with a  background and seven signal functions, 
corresponding to the following states: $\psi^\prime$, $\chi_{c0}$, 
$\chi_{c1}$, $\chi_{c2}$, $\psi^{\prime\prime}$, 
\etacp  and $X(3872)$.  
The resulting fit is shown in  Fig. \ref{psisignal_data}(b), with the yields 
given in Table \ref{final_fits}. 
The $h_c$ charmonium state lies near the \chicone, and it is difficult to distinguish the 
peaks from these two decays. A fit including the $h_c$ yields a number 
of $h_c$ events 
consistent with zero, and a fit performed with free \chicone mass and width  
gives values consistent with a narrow \chicone, 
therefore we have no evidence for $h_c$ production.

\begin{table}[t]
\begin{center}
\caption{
Event yields and absolute branching fractions 
\BR($B^\pm \to K^\pm X_{c\bar c}$) from the fits 
to the $p_K$ spectrum. 
The first error is statistical, the second systematic 
and \BR\ upper limits are given at 90\% CL, 
taking into account the 9.0\% systematic error. 
The last column shows the signal statistical significance $\sigma$,
derived from the fit likelihood assuming 
0 signal events $L(0)$: $\sigma = \sqrt{-2 \log L(0)}$.
For the \etac, both results for absolute and relative measurement, 
and their combination, are reported (see text).} 
\begin{tabular}{|c|c|c|c|c|c|} \hline

Particle &  Yield  & \BR (10$^{-4}$)  & $\sigma$  \\
\hline
\etac           &273$\pm$43   &  8.4$\pm$1.3$\pm$0.8     & 7.3 \\
\etac relative  &             & 10.6$\pm$2.3$\pm$0.4$\pm$0.4 &     \\
\etac combined  &             &  8.7$\pm$1.5             &     \\
\hline
\jpsi           &259$\pm$41   &  8.1$\pm$1.3$\pm$0.7     & 6.9 \\
\chizero        &  9$\pm$21   &  $<$1.8                  & -   \\
\chione         &227$\pm$40   &  8.0$\pm$1.4$\pm$0.7     & 6.0 \\
\chitwo	        &  0$\pm$36   &  $<$2.0                  & -   \\
\etacp          & 98$\pm$52   &  3.4$\pm$1.8$\pm$0.3     & 1.8 \\
\psip	        &139$\pm$44   &  4.9$\pm$1.6$\pm$0.4     & 3.2 \\
\psipp          & 99$\pm$69   &  3.5$\pm$2.5$\pm$0.3     & 1.4 \\
$X(3872)$       & 15$\pm$39   &  $<3.2$                  & -   \\
\hline
\end{tabular}
\label{final_fits}
\end{center}
\end{table}

\begin{table}[t]
\begin{center}
\caption{Summary of systematic errors in percent for absolute and the  
\jpsi : \etac 
relative measurement. 
} 
\begin{tabular}{|c|c|c|} \hline
Source  &  Absolute (\%)& \jpsi\!:\etac  (\%)    \\
\hline\hline
   B counting &  4.5  &  0   \\
\hline
Mass scale  &1&1\\
Background model &3.5 &  1.7   \\
Resolution model &  2.3   &    1.0  \\
\hline
\Kpm reconstruction  &1.3&0\\
\Kpm identification&5&1\\
B mass selection &0.5&0\\
NN1 selection&  2.2    & 2.0 \\
NN2 selection &  3.2   &  1.0  \\ 
\hline
Total &  9.0  & 3.3\\
\hline
\end{tabular}
\label{systematics_tab}
\end{center}
\end{table}

Several sources of systematic error affecting these measurements have been
evaluated. 
The relative errors on absolute measurements are the 
same for all states; many of these cancel partially in relative 
measurements, and all are summarized in Table \ref{systematics_tab}. 
``$B$ counting'' refers to uncertainties in the fit parametrization 
used to determine the number of fully reconstructed $B^\pm_{recon}$. It is one of the largest
errors in absolute measurements, and cancels in ratios. 
The mass scale is verified to a precision of 1.5 MeV/$c$ in $p_K$ by floating the
masses of the   
well-measured  \jpsi\!, \chione and \psip peaks; we assign 
a systematic error corresponding to this shift. 
We also consider variations in the background and
signal model parametrizations, which partially cancel   
in the case of ratios.
Errors in the $K^\pm$ track reconstruction and identification efficiency 
are evaluated by comparing data and MC control samples. 
The systematic error in the NN1 and NN2 selections is evaluated by  
comparing efficiencies and distributions in data and MC, 
and studying efficiency variation with $p_K$.
We verified that the NN2 selection is not 
dependent on visible energy or multiplicity of the recoil part of the $B$ meson decay. 
Adding in quadrature, the total relative error on an absolute measurement is 9.0\%.  
The total is reduced to 3.3\% for the relative measurement of 
\jpsi and \etac\!, and to 5.9\% for states in the high-mass region relative to \jpsi.
For the extraction of relative branching fractions, an additional 4\%
error, labeled (ext) in the following, comes from the present uncertainty of 
\BR$(B^\pm \to K^\pm J/\psi)=(10.0 \pm 0.4) \times 10^{-4}$  \cite{Eidelman:2004wy}.

In the high-mass region, clear signals are found for 
\chione and \psip (with significance 6.0 and 3.2$\sigma$ respectively), 
an excess of events is present 
for \etacp and  \psipp\!\! \cite{Abe:2003zv}, while no signal is found for \chizero\!,  \chitwo  and 
$X(3872)$.
The branching fractions and upper limits are summarized in Table \ref{final_fits}. 

In the low-mass region, our \jpsi measurement is consistent with 
the world average. 
From the \etac and \jpsi yields and the reference branching fraction  
we can derive the result with the relative measurement method $\BR(B^\pm \to K^\pm \eta_c)_{rel}= (10.6\pm 2.3({\rm stat})\pm 0.4({\rm sys}) \pm 0.4 ({\rm ext})) \times 10^{-4}$. 
We combine this result with the absolute measurement of Table \ref{final_fits}, 
taking the correlated errors into account, to obtain $ \BR(B^\pm \to K^\pm \eta_c)=(8.7 \pm 1.5) \times 10^{-4}$.

We obtain from our fits 
the \etac and \etacp masses and widths and   
find $m_{\eta_c} = 2982\pm5$ MeV/$c^2$, $\Gamma_{\eta_c} <43$ MeV  and 
$m_{\eta_c^\prime} = 3639\pm7$ MeV/$c^2$, $\Gamma_{\eta_c^\prime} <23$ MeV,
where the width limits are both at 90\% CL. 

Taking 
\BR$(B^\pm \to K^\pm X(3872)) < 3.2 \times 10^{-4}$, and 
using an average of the Belle \cite{Choi:2003ue} and \babar\  \cite{Aubert:2004ns} measurements of 
\BR($B^\pm \to K^\pm X(3872)$)$\times$\BR($X(3872) \to J/\psi \pi^+
\pi^-$)
we set a lower limit \BR($X(3872) \to J/\psi \pi^+ \pi^-$)$> 4.2 \%$ at 90\% CL. 
This branching fraction, for which there are not yet any predictions, is sensitive to the distribution of charm quarks inside the $X(3872)$. 
A search for charged partners of the $X(3872)$ is performed by examining  
$K^\pm$ recoiling from  a sample of 245.6k reconstructed  
\bz decays. No signal is seen and we find
\BR$(B^0 \to K^\pm\ X(3872)^\mp) < 5\times 10^{-4}$ at 90\% CL.

We combine our \BR($B^\pm \to K^\pm \eta_c$)  
with a previous \babar\  measurement of 
\BR($B^\pm \to K^\pm \eta_c$)$\times$\BR($\eta_c \to K\bar K\pi$)
\cite{Aubert:2004gc} to obtain \BR($\eta_c \to K\bar K\pi$)=$(8.5\pm1.8)\%$, 
significantly improving the precision of the world average 
\cite{Eidelman:2004wy}.
Since this branching fraction is used as a reference for all $\eta_c$ yield measurements, our result will lead to more precise $\eta_c$ partial widths and more stringent comparisons with theoretical models. 
For example, from an average of \BR($J/\psi \to \gamma\eta_c$)$\times$\BR($\eta_c \to K\bar K\pi$) 
measured by Mark-III \cite{Baltrusaitis:1985mr}, DM2 \cite{Bisello:1990re} and BES \cite{Bai:2003tr}, we 
obtain \BR($J/\psi \to \gamma\eta_c$)=(0.79$\pm$0.20)\%, and using the value  
$\Gamma$($\eta_c \to \gamma\gamma$)$\times$\BR($\eta_c \to K\bar K\pi$)=$0.48\pm0.06$ keV \cite{Eidelman:2004wy}
we calculate $\Gamma$($\eta_c \to \gamma\gamma$)=(5.6$\pm$1.4) keV.
Both results are more precise than the world average \cite{Eidelman:2004wy}.
Similarly, we obtain 
\BR(\etacp$\to K\bar K\pi)$=(8$\pm$5)\%
and $\Gamma$(\etacp$\to\gamma\gamma$)=(0.9$\pm$0.5)keV.

In conclusion, a novel technique is used to measure directly the
absolute branching fractions of the various charmonium states $X_{c \bar
c}$ in  two-body decays $B^\pm \to K^\pm X_{c \bar c}$ (Table \ref{final_fits}). 
The results for $X_{c \bar c}=\eta_c, J/\psi, \psi^\prime$ are in agreement 
with previous measurements, and the \etac result significantly improves the
present world average. 
Upper limits are set for \chizero
and \chitwo\!, confirming factorization suppression \cite{Meng:2005}.  
Measurements of $B^\pm \to K^\pm \eta_c^\prime$ and 
$B^\pm \to K^\pm \psi^{\prime\prime}$ branching fractions are reported, 
although with poor significance. 
Upper limits are given for $X(3872)$ and for
production of a possible charged partner in \bz decays.

We are grateful for the excellent luminosity and machine conditions
provided by our \pep2\ colleagues, 
and for the substantial dedicated effort from
the computing organizations that support \babar.
The collaborating institutions wish to thank 
SLAC for its support and kind hospitality. 
This work is supported by
DOE
and NSF (USA),
NSERC (Canada),
IHEP (China),
CEA and
CNRS-IN2P3
(France),
BMBF and DFG
(Germany),
INFN (Italy),
FOM (The Netherlands),
NFR (Norway),
MIST (Russia), and
PPARC (United Kingdom). 
Individuals have received support from CONACyT (Mexico), A.~P.~Sloan Foundation, 
Research Corporation,
and Alexander von Humboldt Foundation.

\end{document}